# Domain Control by Adjusting Anisotropic Stress in Pyrochlore Oxide $Cd_2Re_2O_7$

Satoshi Tajima[1], Daigorou Hirai[1], Yuto Kinoshita[1], Masashi Tokunaga[1], Kazuto Akiba[2], Tatsuo C. Kobayashi[2], Hishiro T. Hirose[3], and Zenji Hiroi[1*]

[1]*Institute for Solid State Physics, University of Tokyo, Kashiwa, Chiba 277-8581, Japan*
[2]*Graduate School of Natural Science and Technology, Okayama University, Okayama 700-8530, Japan*
[3]*National Institute for Materials Science, Tsukuba, Ibaraki 305-0003, Japan*
*E-mail: hiroi@issp.u-tokyo.ac.jp

The 5d pyrochlore oxide $Cd_2Re_2O_7$ exhibits successive phase transitions from a cubic pyrochlore structure (phase I) to a tetragonal structure without inversion symmetry below $T_{s1}$ of ~200 K (phase II) and further to another noncentrosymmetric tetragonal structure below $T_{s2}$ of ~120 K (phase III). The two low-temperature phases may be characterized by odd-parity multipolar orders induced by the Fermi liquid instability of the spin–orbit-coupled metal. To control the tetragonal domains generated by the transitions and to obtain a single-domain crystal for the measurements of anisotropic properties, we prepared single crystals with the (0 0 1) surface and applied biaxial and uniaxial stresses along the plane. Polarizing optical microscopy observations revealed that inducing a small strain of approximately 0.05% could flip the twin domains ferroelastically in a reversible fashion at low temperatures, which evidences that the tetragonal deformation switches at $T_{s2}$ between $c > a$ for phase II and $c < a$ for phase III. Resistivity measurements using single-domain crystals under uniaxial stress showed that the anisotropy was maximum at around $T_{s2}$ and turned over across $T_{s2}$: resistivity along the $c$ axis is larger (smaller) than that along the $a$ axis by ~25% for phase II (III) at around $T_{s2}$. These large anisotropies probably originate from spin-dependent scattering in the spin-split Fermi surfaces of the cluster electric toroidal quadrupolar phases of $Cd_2Re_2O_7$.

## 1. Introduction

The pyrochlore oxide $Cd_2Re_2O_7$ (CRO) has attracted attention as a candidate for the spin–orbit-coupled metal (SOCM).[1,2] SOCM is a metallic system characterized by a centrosymmetric crystal structure and conduction electrons with strong spin–orbit interactions. It has a specific Fermi liquid instability that causes spontaneous inversion symmetry breaking (ISB) leading to unusual odd-parity orders such as multipolar and gyrotropic orders.[1] In such odd-parity orders of itinerant electrons, one expects an intriguing interplay between spin and charge degrees of freedom, because antisymmetric spin–orbit coupling (ASOC) activated by ISB gives rise to a sizable spin splitting in the original spin-degenerate Fermi surface (FS); for example, unconventional magneto-current effects are predicted for an odd-parity multipolar phase.[3-5] Since CRO is an itinerant $5d^2$ electron system with a large spin–orbit interaction and crystallizes in a cubic pyrochlore structure with inversion symmetry at room temperature, it has been assumed to be a candidate SOCM. In fact, it exhibits an ISB transition accompanied by large changes in electronic properties.[2]

CRO undergoes two structural transitions at $T_{s1}$ of ~200 K and $T_{s2}$ of ~120 K.[6] Phase I above $T_{s1}$ crystallizes in a cubic structure of the space group $Fd\bar{3}m$, which is composed of regular tetrahedra of Re atoms linked by their vertices to form a pyrochlore lattice.[7] In phase II below $T_{s1}$, the space inversion center is lost together with the threefold rotation axis and a tetragonal structure of the space group $I\bar{4}m2$ is formed. Since this second-order ISB transition is accompanied by large changes in physical properties, particularly an approximately 50% decrease in the density of states, it probably originates from the Fermi liquid instability of SOCM.[1,2] Another tetragonal phase III of the space group $I4_122$ without inversion symmetry appears below $T_{s2}$, at which relatively small changes in physical properties are observed. According to the SOCM scenario, these two low-temperature (LT) phases should have odd-parity multipolar orders.[1,2]

In phase III, CRO is rendered superconducting below $T_c$ = 1.0 K.[8-10] Although an exotic superconducting state was expected in the absence of inversion symmetry,[11-13] all the results of experiments have revealed that CRO is a conventional weak-coupling superconductor with an isotropic gap.[10,14-19] Under a high pressure, however, an unconventional face of superconductivity seems to emerge near the critical pressure $P_c$ of ~4 GPa at which the ISB transition disappears.[20,21] Moreover, the results of recent Re nuclear quadrupole resonance (NQR) experiments suggested an exotic superconductivity induced by parity fluctuations above $P_c$.[22]

To gain insight into the physics of SOCM in CRO, it is crucial to clarify the characteristic of the two LT phases in detail and identify the corresponding odd-parity multipolar orders. On the basis of exact symmetry arguments and possible interpretations of the accumulated experimental data, nonmagnetic electric toroidal quadrupole (ETQ) orders of the $x^2 - y^2$ and $3z^2 - r^2$ types,[3,5] an electric dotriacontapole order with a magnetic quadrupole component,[4] and a combination of odd-parity quadrupolar and even-parity magnetic octupolar orders[23-25] have been theoretically proposed. The former ETQ scenario is based on the successive symmetry reductions mentioned above and assumes an $E_u$ order parameter (OP), whereas the latter two models rely on the nonlinear optical measurements developed by Harter et al., the results of which suggested $T_{2u}$ and $T_{1g}$ OPs, indicating lower-symmetry structures for phase II.[23,24,26] Phase III was not considered because the second transition at $T_{s2}$ was not observed in the nonlinear optics experiments. On the other hand, recent magnetic torque measurements showed the possibility of parity mixing in the primary OPs: an even-parity $E_g$ OP plus an odd-parity OP such as $E_u$.[27]

Critical information for solving the above controversy and identifying the true OP(s) will be obtained by experiments that can reveal the types of spin splitting in the LT phases. One expects unconventional magneto-current effects, such as the Edelstein or rectifying effects, which reflect the symmetries of spin-momentum locking induced by ASOC.[4,5] Moreover, anisotropies in physical properties, which must be dominated by spin-split FSs, are to be examined. Nevertheless, such experiments have been difficult thus far because of complexity generated by the formation of tetragonal domains at the cubic-to-tetragonal transition, which was actually observed by polarizing optical microscopy (POM);[26,28] any anisotropy may be averaged over. In previous de Haas–van Alphen experiments, the FSs of phase III were deduced by taking into account the contributions from all domains.[29] However, it is desirable to carry out more experiments using a single-domain crystal of CRO in order to obtain reliable data on the LT phases; this will become possible under anisotropic stress.

The size of lamella domains in the twins of CRO is large, up to 500 μm in thickness, which reflects a tiny tetragonal distortion of 0.05% estimated in the single-crystal X-ray diffraction study[30] or 0.10% in powder neutron diffraction experiments.[31] Both experiments showed tetragonal deformations with $c < a$ in a pseudocubic notation down to 10 K; herein, we use the pseudocubic notation unless otherwise stated. In contrast, it was predicted on the basis of the Landau theory for phase transitions[32,33] that the tetragonality should flip at $T_{s2}$,[34] although the magnitude relation between $c$ and $a$, which is determined by the Landau parameter $\lambda_2$ containing the elastic stiffness constants,[34] had not yet been known. Experimentally, such flipping was suggested from previous POM observations of the (1 1 1) surface showing a marked change in domain pattern at $T_{s2}$.[28] The tetragonality and its possible flip at $T_{s2}$ will be unequivocally determined by examining the responses of domains to anisotropic stresses.

In this study, we prepared single crystals with the (0 0 1) surface and applied biaxial and uniaxial stresses along the plane using piezoelectric (PE) elements. The POM observations showed that the twin domains of CRO were well controlled reversibly by stresses that caused strains of ±0.1% at maximum. Under biaxial stress, a single-domain crystal with the $c$ axis perpendicular to the surface was obtained under compression in phase II and expansion in phase III. Under uniaxial stress, on the other hand, a single-domain crystal with the $c$ axis along the stress direction was obtained under expansion in phase II and compression in phase III. From these responses, it is unequivocally concluded that the tetragonality flips at $T_{s2}$ between $c > a$ for phase II and $c < a$ for phase III. In addition, anisotropy in resistivity was measured using single-domain crystals under uniaxial stress, and it was found that significant anisotropies as large as 25% also flipped at $T_{s2}$. The origin of the observed anisotropies in the crystal structure and resistivity is discussed on the basis of the cluster-ETQ model for phases II and III.

## 2. Experiments

The crystals of CRO thus far studied have an octahedral shape with (1 1 1) facets. However, they are not suitable for the manipulation of the tetragonal domains; pressure must be applied along the [1 0 0] direction. In addition, POM observations on a natural (0 0 1) surface can give more straightforward information on the formation of twin domains than those on a (111) surface; previous POM observations gave no domain contrast on a polished crystal surface.[28] We were successful in preparing truncated-octahedral crystals with (0 0 1) facets, as shown in Fig. 1(a) in two steps via a modified synthesis route: first, octahedral crystals were grown by a chemical vapor transport (CVT) method starting from a mixture of CdO:ReO$_3$:Re = 1.44:1:0.2 at a temperature gradient of 1000–1100 K for 100 h, and second, recrystallization was carried out by a CVT method at a temperature gradient of 900–960 K for 300 h. Occasionally, we obtained large crystals with a 1 mm edge of the (0 0 1) surface. For stress experiments, several thin plates of ~0.1 mm thickness were hewed out along the (0 0 1) surface from crystals grown under the same conditions.

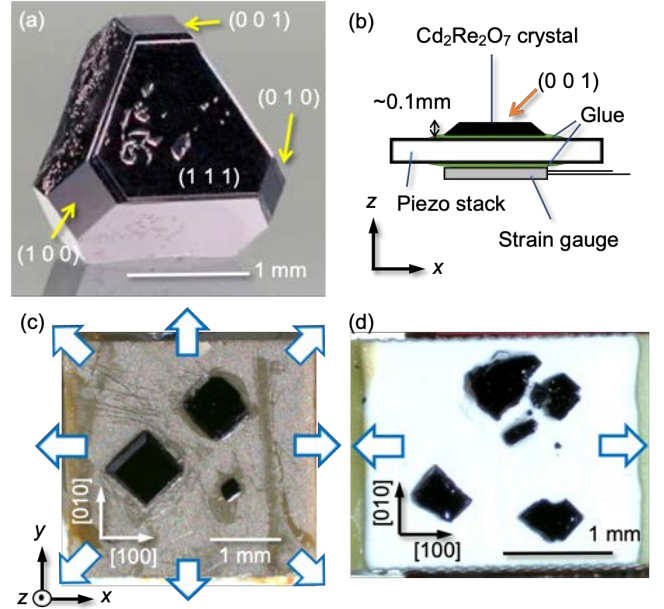

**Fig. 1**. (Color online) (a) Typical CRO crystal of a truncated octahedral shape with natural {1 0 0} and {1 1 1} surfaces. (b) Schematic representation of a specimen for polarizing optical microscopy observations and resistivity measurements under biaxial and uniaxial stresses. A CRO crystal with a natural (0 0 1) surface was cut into a piece of ~100 μm thickness and adhered to a PE element using epoxy adhesive (Araldite Standard). On the back of the PE element, a strain gauge was attached to monitor the deformation of the element. (c) Setup for biaxial stress experiments. Three crystals were put on a PE element that expands or shrinks along the plane, as marked by the large open arrows. (d) Setup for uniaxial stress experiments. The uniaxial stress was applied along the horizontal direction ($x$) parallel to the crystallographic [1 0 0] direction in the pseudocubic notation.

Tunable in-plane biaxial and uniaxial stresses were applied to CRO single crystals using PE elements, following the methods used in previous studies.[35-37] The PE elements used here were multilayer actuators composed of stacks of ferroelectric materials; PL055.31 (PI Ceramic) for biaxial stress and PSt 150/3.5×3.5/7 Cryo 1 (Keystone International Inc.) for uniaxial stress. A few plate crystals were glued onto the side wall of a PE stack on their polished faces [Figs. 1(b)–(d)], so that they were forced to deform together with the PE element. The deformation of the PE element was controlled by applying a known voltage, and the induced strain (the fractional change of length, $\varepsilon = \Delta L/L$) was monitored via a strain gauge glued onto the back of the PE element [Fig. 1(b)]; the strain gauges used were KFL-2-120-C1-11 and KFL-05-120-C1-11 (Kyowa Electronic Instruments Co., Ltd.) for biaxial and uniaxial stresses, respectively. For biaxial stress, the PE element deformed along the $z$ direction perpendicular to the crystal plane, which caused an in-plane deformation to be transmitted to the crystals in accordance with the Poisson ratio. On the other hand, for uniaxial stress, the PE element deformed along the $x$ axis in the crystal plane, accompanied by an additional stress of the opposite sign along the $y$ axis. In either case, compressive and expansive strains of approximately ±0.1% at maximum were induced at voltages of ±100–200 V.

POM was employed to observe the formation of domains and their responses to applied stresses. The observations were carried out using an optical microscope (Olympus, BXFM) in the reflection mode.[28] Images were recorded using a monochrome CCD camera (ST-402ME, SBIG). A pristine (0 0 1) crystal surface was observed in a crossed–Nicols setup. The samples shown in Figs. 1(c) and 1(d) were attached to sapphire plates with silver paste and cooled slowly in a GM refrigerator. Two series of observations were carried out at 150 and 90 K while scanning the voltage applied to the PE elements.

Electrical resistivity was measured under uniaxial stress by a standard four-probe method with four terminals approximately aligned along the $x$ [Fig. 7(b)] or $y$ direction [Fig. 8(b)] of two crystals obtained under the same preparation conditions; the former crystal was also used for POM observations, which confirmed the tuning of domains by applying stress. A low-resistivity contact was achieved via the indium metal between the sample and silver paste connecting to the gold wire probes. The sample was set to an Oxford Optical Cryostat and was cooled to 4 K. Isothermal measurements were performed at 210 (phase I), 150 (II), and 90 K (III), at which three loops were recorded as a function of the voltage applied to the PE elements. The temperature dependence of resistivity was measured with the current along the $x$ direction upon cooling at 2 K/min at a voltage of 200 V ($\varepsilon = 0.07$–0.08%) or –65 V ($\varepsilon = -0.11$– –0.15%).

The electronic structures of CRO were calculated on the basis of the density functional theory[38] using the Quantum ESPRESSO package.[39] The fully relativistic projector augmented-wave method[40] with a plane-wave cutoff energy of 54 Ry, a 10 × 10 × 10 k-point mesh, and the Perdew–Burke–Ernzehof exchange potential[41] for the self-consistent field calculation was employed. The FSs and their spin polarizations were calculated using a 24-orbital tight-binding model based on maximally localized Wannier functions constructed by the Wannier90 program.[42] The calculations were based on the structural data given by Huang et al.[43] In addition, the anisotropy in electronic conductivity was evaluated using the Boltzmann transport equation in the constant relaxation time approximation, as implemented in the Wannier90 program.[44]

## 3. Results
### 3.1 Polarizing optical microscopy observations
#### 3.1.1 Domain formation upon cooling

We first show how the temperature evolution of twin domains is observed in the POM on a (0 0 1) crystal surface in the absence of stress. As shown in Fig. 2, a uniform contrast at 205 K above $T_{s1}$ changes to a contrast with striation along [1 1 0] at 195 K, indicating the formation of twins made of lamella domains in the tetragonal structure with optical anisotropy. Upon further cooling to 130 K, the domain contrast is enhanced and the thickness of lamellas increases to ~100 μm at maximum. Then, at 100 K below $T_{s2}$, other types of striation along [1 –1 0] and [1 0 0] appear. Similar changes showing domain flipping were recorded in the previous POM observations on the (1 1 1) surface and ascribed to the switching of the tetragonality at $T_{s2}$.[28]

There is another type of grey–white contrast in Fig. 2 made of not straight lines as in the twins but some curved lines that form a circle [left bottom in Figs. 2(b)–(d)]. Since this contrast appears below $T_{s1}$, it must be caused by the phase transition. However, it is not affected by the temperature evolution of the twin domains or by their marked change at $T_{s2}$. This additional contrast was observed on a few but not all of the crystals examined in our POM experiments, and its origin is not clear.

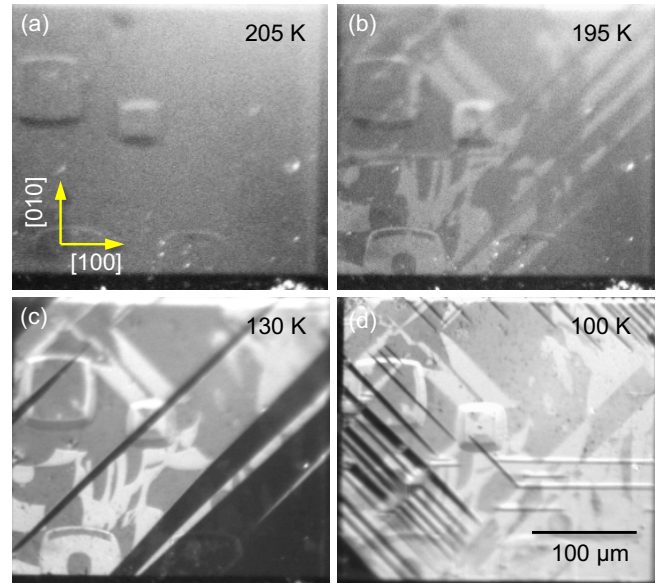

**Fig. 2.** (Color online) Temperature evolution of twin domains observed by polarizing optical microscopy on the (0 0 1) surface at (a) 205, (b) 195, (c) 130, and (d) 100 K. No stress was applied during observations. The two square terraces on the surface at the upper left of all images are the natural markers for identical areas observed upon cooling. The striations with intervals of 10–100 μm along [1 1 0], [1 –1 0], and [1 0 0] are attributed to twin domain boundaries that appear below $T_{s1}$ and change their fractions across $T_{s2}$. The origin of the additional grey–white contrast with meandering boundaries, which appears below $T_{s1}$ and is not affected at $T_{s2}$, is not known.

*3.1.2 Domain types and boundaries*

The geometrical relations between the cubic unit cell of phase I and the tetragonal ones of phases II and III are schematically depicted in Fig. 3(a). The transitions between them involve the generation of three types of domain with their surviving fourfold rotation axes ($c$ axis) along $x$, $y$, and $z$; these domains are named X, Y, and Z, respectively. Two of the domains join coherently at the invariant interface planes of the {1 1 0} type to form a transformation twin.

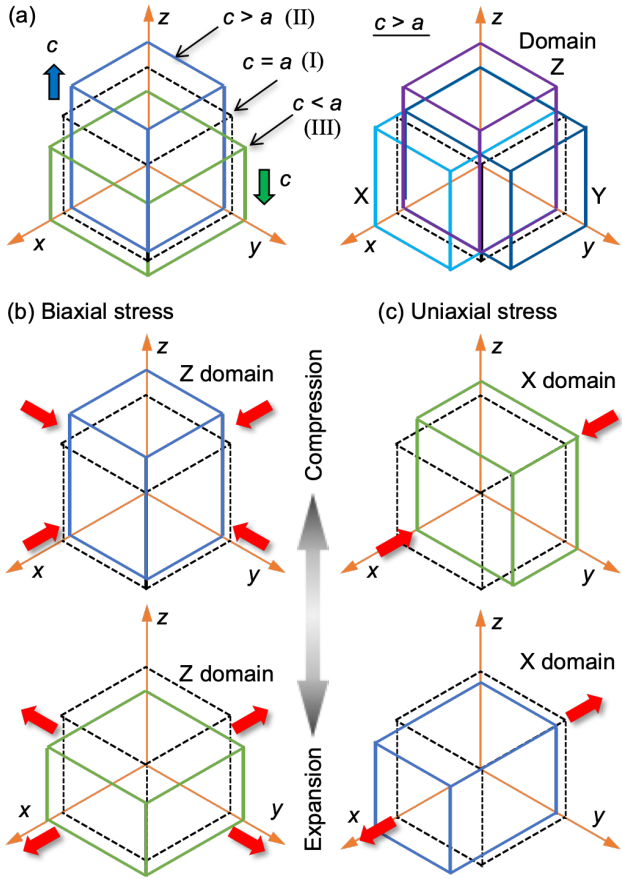

**Fig. 3.** (Color online) (a) Schematic representations for two tetragonal deformations realized in CRO below $T_{s1}$ and $T_{s2}$ (left) and three tetragonal domains with the fourfold rotation axis ($c$ axis) along $x$ (domain X), $y$ (Y), and $z$ (Z) in the case of $c > a$, where the lattice constants assume a pseudocubic unit cell. The two deformations from the cubic unit cell of phase I (broken lines) with $c > a$ (blue lines) and $c < a$ (green lines) correspond to the cases for phases II and III, respectively. (b, c) Possible selections of tetragonal domains into single domains under (b) biaxial and (c) uniaxial stresses; in each figure, the red arrows represent the directions of applied stresses, which vary from compression (top) to expansion (bottom).

A typical POM image including all possible domains is shown in Fig. 4, which was recorded at 150 K for phase II without stress. Lamella domains with thicknesses of several tens of µm and with three types of contrast (dark, grey, and bright) are discernible. They form three types of striation with different orientations: diagonal along (1 1 0) or (1 –1 0) between domains X and Y, horizontal along (0 1 1) or (0 –1 1) between domains Y and Z, and vertical along (1 0 1) or (1 0 –1) between domains Z and X. Note that the former X–Y boundaries are exactly perpendicular to the image plane, while the latter two boundaries are inclined by approximately 45° from the image plane. Note also that the $c$ axes are slightly inclined from the original cubic Cartesian axes so as to keep the corresponding interface planes invariant. From these identifications of the twin boundaries, it is uniquely determined that X, Y, and Z domains have bright, grey, and dark contrasts, respectively. This is reasonable, because only the Z domain has no optical anisotropy; the difference between the X and Y domains originates from the relative orientations of the incident polarizer and output analyzer used in the experimental setup. There are fine striations, which are due to secondary twinning, separated by 5–10 µm inside each domain: for example, the Y–Z twin domains at the right side of Fig. 4 contain fine X domains, giving vertical boundaries inside the Z domains and diagonal ones inside the Y domains.

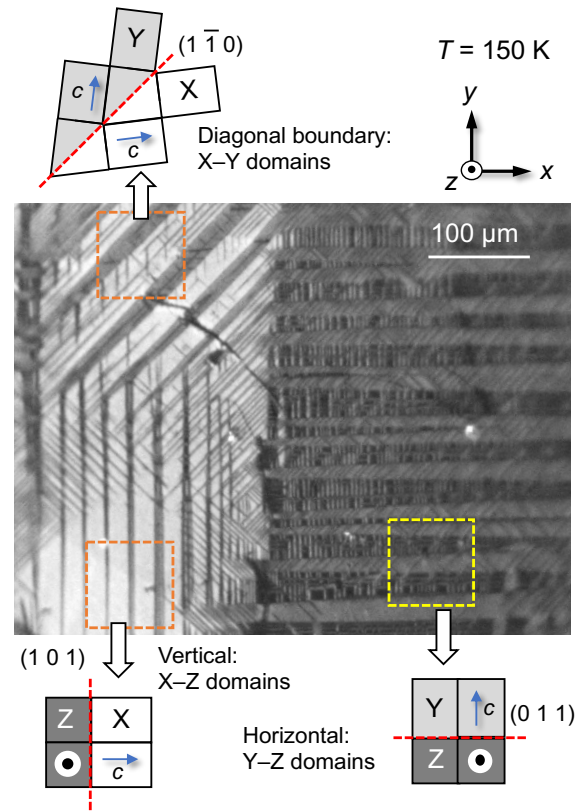

**Fig. 4.** (Color online) Typical POM image recorded at 150 K in the absence of stress. The diagonal stripes at the upper left correspond to the twinning of X and Y domains, which are exactly parallel to the (1 1 0) invariant plane, as schematically depicted above the image; the tetragonal distortion is exaggerated for clarity. The nearly vertical and horizontal stripes at the lower left and right show the Z–X and Y–Z twins, respectively. The bright, grey, and dark contrasts correspond to the X, Y, and Z domains, respectively. Inside each stripe, there are fine striations due to the secondary twining of the third domain; i.e., the X domain in the Y–Z twin. The Cartesian axis of the original cubic cell is shown at the upper right; the $c$ axis of the tetragonal unit cell in each domain is inclined from the Cartesian axis by a minimal angle; those in the X–Y twin are inclined toward the [1 1 0] direction, and those in the Y–Z and X–Z twins toward the [0 1 1] and [1 0 1] directions, respectively.

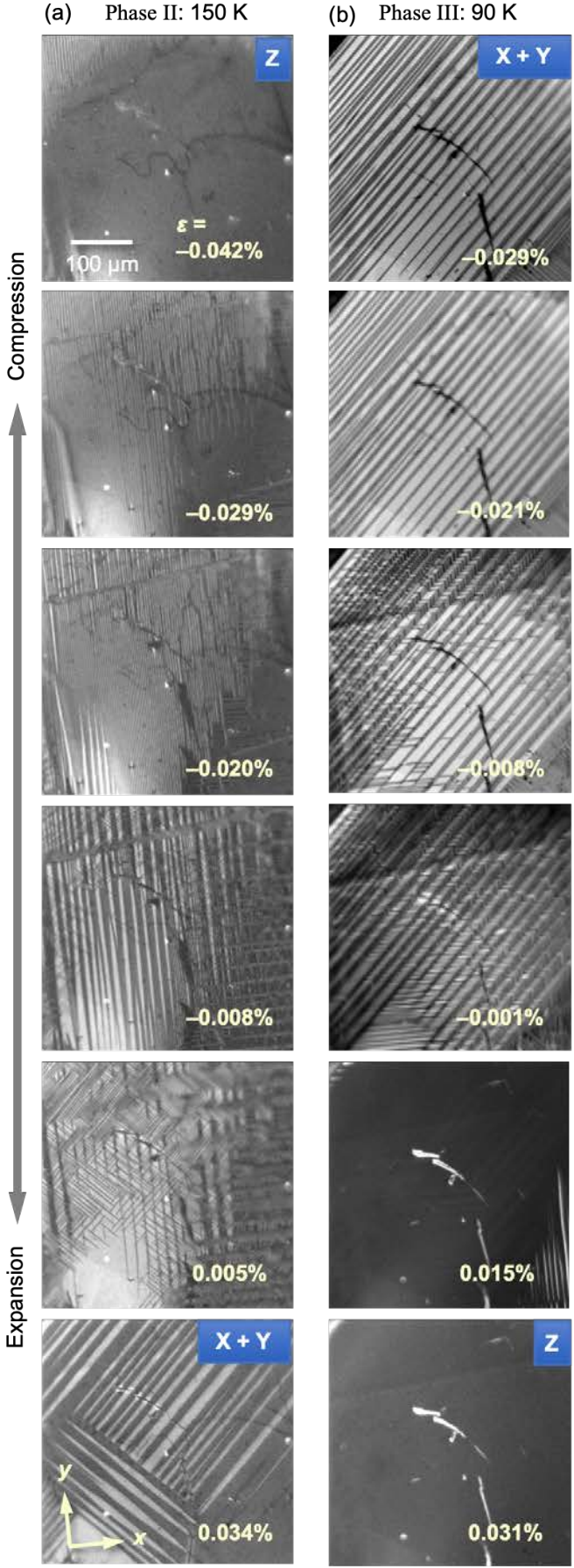

**Fig. 5.** (Color online) Evolution of domain contrast under biaxial stress at (a) 150 and (b) 90 K. The percentage value in each image gives the magnitude of induced strain measured using the strain gauge. The position of scars [e.g., bright lines in the bottom image of (b)] in each image guarantees that the same area of the crystal is observed during the stress scan and temperature change. The crystal is biaxially compressed to experience negative strain toward the top and is under tension with positive strain down to the bottom. A single-Z-domain crystal with the $c$ axis perpendicular to the plane of the paper was attained under compression at 150 K in phase II and under expansion at 90 K in phase III, as indicated by the uniform contrast with dark color. On the other hand, X–Y twins with the $c$ axes nearly lying along the plane of the paper were attained under expansion in phase II and compression in phase III, as indicated by the diagonal striations without vertical or horizontal striations. All the variations were completely reversible against the stress scan.

*3.1.3 Response of domains to biaxial stress*

Now we describe how the domains in CRO responded to biaxial stress. Figure 5 shows two series of POM images recorded first at 150 K for phase II with changing applied stress and then at 90 K for phase III. At 150 K, the images taken at small strains are similar to that in Fig. 4, indicating mixtures of three types of domain. With increasing compressive strain, the striations fade away and completely disappear at $\varepsilon = -0.042\%$, resulting in a dark uniform image, which indicates that a single-Z-domain crystal has been obtained. On the other hand, when an expansive strain of 0.034% is induced, the image is made of only diagonal boundaries of X–Y twins. At 90 K, in contrast, a single Z domain with a dark contrast is obtained at an expansive strain of 0.031%, while X–Y twins appear at $\varepsilon < -0.02\%$. Therefore, the responses to strain are reversed in the two experiments and probably between the two phases. These variations of the domain distribution as a function of strain are reproducible and reversible.

The observed domain responses are straightforwardly interpreted by considering the stability of tetragonal domains under biaxial stress. As depicted in Fig. 3(b), for a tetragonal distortion of $c > a$, a single Z domain and double X–Y domains are selected under compressive and expansive stresses, respectively, and vice versa for $c < a$. Since the former corresponds to the observations at 150 K and the latter to those at 90 K, it is concluded that phases II and III have tetragonal distortions of $c > a$ and $c < a$, respectively (Table 1). Thus, a switching of tetragonality indeed occurs at $T_{s2}$, as suggested by previous POM observations on (1 1 1) surfaces[28] and from the Landau theory;[34] the Landau parameter $\lambda_2$ is determined to be positive.

*3.1.4 Domain control by uniaxial stress*

The obtained single-Z-domain crystal with the $c$ axis perpendicular to the crystal surface under biaxial stress was not suitable for in-plane resistivity measurements to determine the electronic anisotropy. Thus, we carried out similar experiments under uniaxial stress. What one expects under uniaxial stress along the $x$ direction is illustrated in Fig. 3(c): a single-X-domain crystal should occur with an expansive strain for phase II with $c > a$ and a compressive strain for phase III with $c < a$.

Figure 6 shows two sets of POM images under uniaxial stress. Horizontal striations indicating Y–Z twins are observed under compression at 150 K and under expansion at 90 K, while uniform contrasts indicating single X domains are observed under expansion at 150 K and under compression at 90 K, which are exactly as expected above. The magnitudes of strain to achieve double and single domains were approximately 0.05%. Similarly to the cases of biaxial stress, these changes in domain patterns were reversible, indicating that the domain boundaries were mobile at low temperatures.

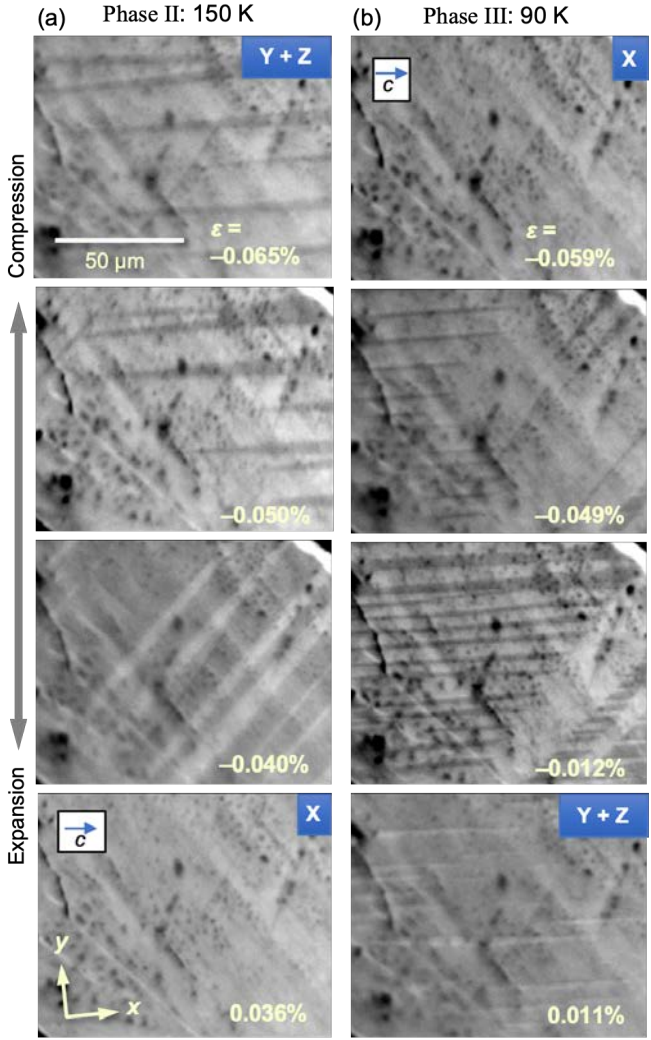

**Fig. 6.** (Color online) Evolution of the domain pattern under uniaxial stress at (a) 150 and (b) 90 K. The crystal was uniaxially compressed along the horizontal direction ($x$) toward the top and was under tension to the bottom. Single-X-domain crystals with the $c$ axes along the horizontal direction were attained under tension in (a) and compression in (b), while Y + Z twins were attained under compression in (a) and tension in (b). The overall contrasts were weak compared with those in Fig. 5 probably because of the poor crystallinity or surface of the examined crystal.

**Table 1.** Anisotropies in the lattice, magnetic susceptibility, and resistivity of CRO.

| Property \ Phase | | I | II | III | Reference |
|---|---|---|---|---|---|
| Lattice | Relation[*a] | $c = a$ | $c > a$ | $c < a$ | This study |
| | Anisotropy[*b] | 0 | ~0[*c] | ~0[*c] | Refs. 30, 31, 43 |
| Magnetic susceptibility | Relation | $\chi_c = \chi_a$ | $\chi_c < \chi_a$ | $\chi_c > \chi_a$ | Ref. 28 |
| | Anisotropy[*d] | 0 | −11%[*e] | 11%[*f] | |
| Resistivity | Relation | $\rho_c = \rho_a$ | $\rho_c > \rho_a$ | $\rho_c < \rho_a$ | This study |
| | Anisotropy[*g] | 0 | 25% | −25% | |

[*a]In the pseudocubic notation. [*b]The anisotropy is defined as $2(c − a)/(c + a)$. [*c]The magnitude of the tetragonal distortion may be as small as 0.05%,[30)] 0.10%[31)] or 0.01%.[43)] [*d]The anisotropy is defined as $2(\chi_c − \chi_a)/(\chi_c + \chi_a)$. [*e]$T = 120$ K. [*f]$T = 4$ K. [*g]The anisotropy is defined as $2(\rho_c − \rho_a)/(\rho_c + \rho_a)$ and the values were obtained at around $T_{s2}$.

*3.2 Resistivity measurements under uniaxial stress*
*3.2.1 Domain switching*

Resistivities along the $x$ and $y$ directions were measured under uniaxial stress along $x$ at $T = 210$, 150, and 90 K. Figure 7 shows the $R_x$ data as a function of strain. There is no change at 210 K for phase I, whereas either of the LT data exhibits a large hysteresis loop with an increase at expansion ($\varepsilon > 0$) and a decrease at compression ($\varepsilon < 0$): the difference reaches 17–18%. Note that the saturation of the hysteresis loops occurs above $\varepsilon$ of ~0.05% and below −0.02% from the loop center in either case. Taking into account the corresponding POM images, these saturations are found to be due to the termination of domain changing into single or double domains; this also means that the substantial strain dependence of resistivity in CRO is negligible in the present range. On the other hand, the large open loop at small strains is ascribed to the coexistence of the three types of domain, as is the cases for ferroelectrics and ferromagnetism. The domain distribution is determined to reduce the total energy in strain and depends on the initial state. The reason why the hysteresis loop is shifted to the left from the origin is that the crystal already suffered from an additional expansive stress probably owing to the mismatch in thermal expansion between the crystal and the PE element; in fact, the center of the loop shifts further to the left upon cooling. The reason for the gradual increase in $R_x$ at $\varepsilon > 0$ at 90 K is not known, but may be related to a misalignment in the crystal setup.

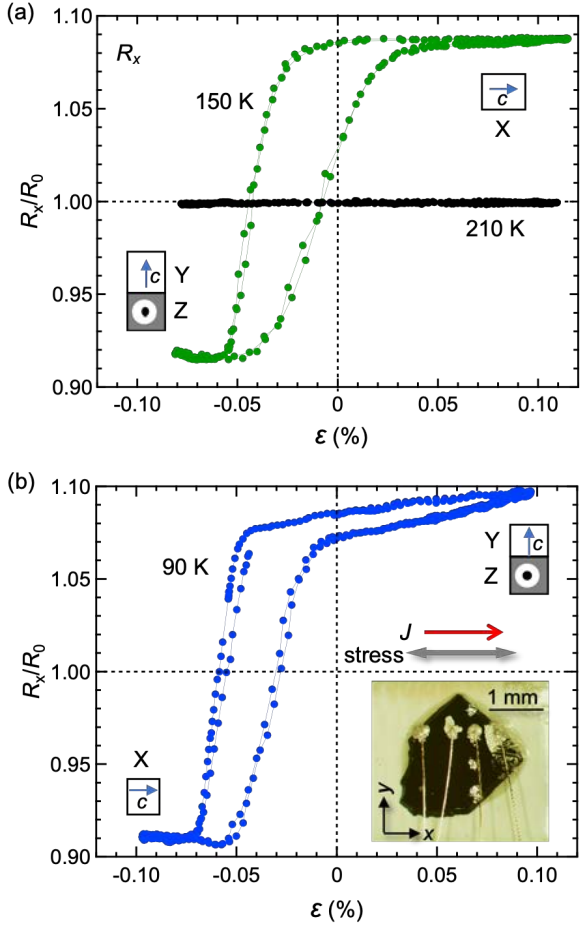

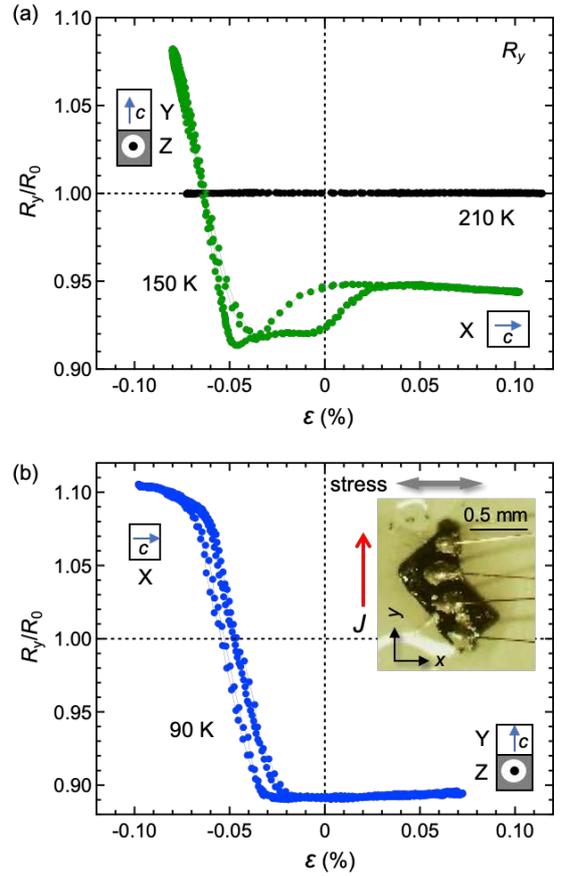

90 K. At 150 K, $R_a$ was measured from a single X domain under expansion, while both Y and Z domains contributed to $R_y$ with $R_c$ and $R_a$ components under compression, respectively; the latter is not an average of $R_c$ and $R_a$, because the interface between Y and Z domains is inclined by about 45° from the electric current direction. At 90 K, $R_a$ was measured from a single X domain under compression, while Y and Z domains contributed to $R_y$ with $R_c$ and $R_a$ components under expansion, respectively. Thus, the anisotropy in $R_y$ is consistent with the fact that $R_c > R_a$ for phase II and $R_c < R_a$ for phase III from the $R_x$ measurements. Note that the 150 K curve shows complex behavior at small strains, which is difficult to explain by the simple selections of domains. The reason for this may be a misalignment causing an in-plane rotation of the crystal, generating Y domains having a larger $R_c$ contribution along the $y$ direction in addition to the ideal X domains under expansion.

**Fig. 7.** (Color online) Variations in $R_x$ as a function of strain $\varepsilon$ along the $x$ direction at (a) 210 and 150 K and (b) 90 K. The photograph in (b) shows the experimental setup with four probes to measure resistance along the $x$ direction, which is parallel to the direction of the uniaxial stress. The resistance was normalized to the value at the center of the hysteresis loop at each temperature. Zero strain means that zero voltage was applied to the PE element. The hysteresis loop shifts to the left from the origin because of an additional expansive stress exerted by the mismatch in thermal expansion between the crystal and the PE element. The insets in (a) and (b) depict the corresponding selections and orientations of domains at the maximum strain.

How the domains switch under uniaxial stress is schematically depicted in Fig. 7. Taking into account these domain selections, it is apparent that, at 150 K for phase II, the $c$-axis component of resistance $R_c$ was measured from a single X domain at the maximum expansion, while the $a$-axis component $R_a$ was obtained from Y–Z domains under the maximum compression. At 90 K for phase III, $R_c$ and $R_a$ were measured under the opposite conditions of maximum compression and expansion, respectively. Thus, we conclude that there are significant anisotropies in resistivity in each phase and, interestingly, that they flip between $R_c > R_a$ for phase II and $R_c < R_a$ for phase III (Table 1).

The resistance $R_y$ under uniaxial stress along $x$ was measured to confirm the above results for $R_x$. As shown in Fig. 8, $R_y$ actually behaves in an inverse fashion to $R_x$, as expected for each phase. No strain dependence was observed at 210 K, and hysteresis loops with magnitudes of ~20% and opposite responses to those shown in Fig. 7 were recorded at 150 and

**Fig. 8.** (Color online) Variations in the resistance $R_y$ measured along the $y$ direction as a function of strain along the $x$ direction at (a) 210 and 150 K and (b) 90 K. The resistance was normalized to the average of the maximum and minimum values at each temperature. The photograph in (b) shows the experimental setup with four probes to measure resistance along the $y$ direction. The insets in (a) and (b) depict the corresponding domain selections and orientations in an ideal situation. The complex behavior at small strains in (a) may be due to a misalignment of the crystal in the experimental setup.

*3.2.2 Anisotropy in resistivity*

The temperature dependences of resistivity obtained from the $R_x$ measurements under the maximum expansion with $\varepsilon = 0.07$–$0.08\%$ and the maximum compression with $\varepsilon = -0.11$–

−0.15% are shown in Fig. 9. The two curves are located above or below another curve obtained from a freestanding CRO crystal having multiple domains, suggesting that the third curve is their average; this also suggests a minor contribution from scattering at domain boundaries. Note that the expansion and compression curves correspond to $R_c$ and $R_a$ in phase II and to $R_a$ and $R_c$ in phase III, respectively. The difference between $R_c$ and $R_a$ gradually increases below $T_{s1}$ and suddenly flips at $T_{s2}$. The two curves happen to be smoothly connected at $T_{s2}$, but there are small changes in the slopes for the expansion curve, as shown in the inset of Fig. 9.

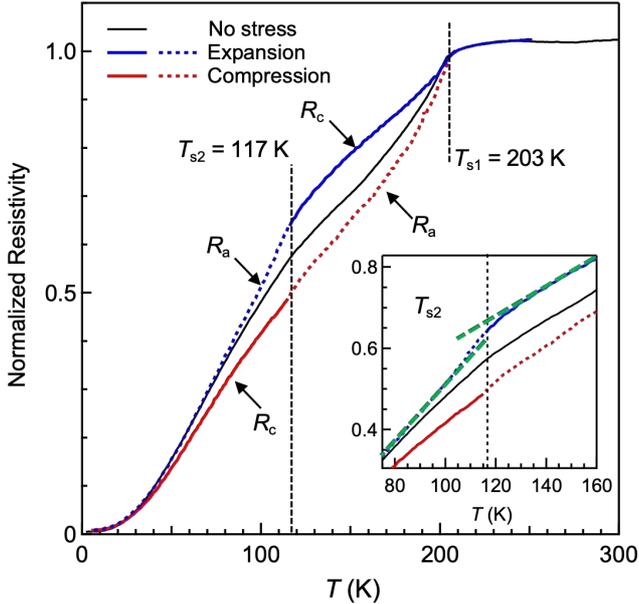

**Fig. 9.** (Color online) Temperature dependences of $R_x$ measured upon cooling below 250 and 210 K to 4 K under uniaxial stresses along the $x$ direction at the maximum expansion (blue line) and the maximum compression (red line), respectively. The magnitudes are normalized to the resistivity at 210 K measured on a freestanding crystal (black line). Taking into account the observed domain alignments, the upper expansive and lower compressive curves are found to correspond to $R_c$ (solid lines) and $R_a$ (broken lines) above $T_{s2}$ in phase II and to $R_a$ and $R_c$ below $T_{s2}$ in phase III, respectively. The inset shows an enlargement of the variations around the $T_{s2}$ transition, in which the green broken lines are guides for the eyes.

The anisotropy in resistivity, which is defined as $2(R_c - R_a)/(R_c + R_a)$, is plotted in Fig. 10. It increases gradually below $T_{s1}$, reflecting the second-order characteristic, and reaches 25% immediately above $T_{s2}$. Then, it flips to −25% and gradually decreases to −10% at the lowest temperature. Therefore, the anisotropy in the electronic states develops as a result of the ISB transition and changes markedly at the $T_{s2}$ transition. Although the $T_{s2}$ transition has been considered subtle compared with the $T_{s1}$ transition, there must be a crucial change in electronic state at $T_{s2}$, as already suggested in previous studies.[2,45,46]

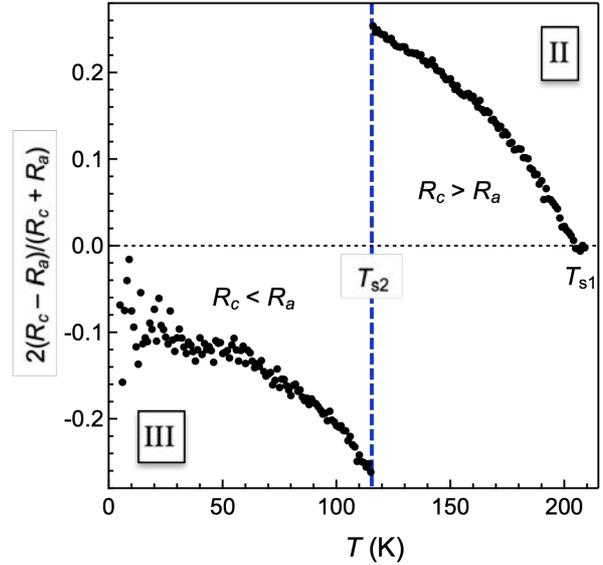

**Fig. 10.** (Color online) Temperature dependence of the anisotropy in resistivity. Upon cooling, the anisotropy increases gradually below $T_{s1}$ toward $R_c > R_a$ and reaches 25% immediately above $T_{s2}$. Then, it jumps to −25% immediately below $T_{s2}$ and gradually decreases to − 10% at $R_c < R_a$.

## 4. Discussion
### 4.1 Anisotropic stress effects and lattice distortions
#### 4.1.1 Anisotropic stress effects in general

Recent progress in anisotropic stress techniques using PE elements has made it possible to induce a strain as large as 1% in a crystal, which leads to the discovery of various interesting phenomena.[47,48] Particularly in the case of symmetry breaking in the electronic state coupled with symmetry lowering in the crystal structure, an extraordinary response emerges even at a small strain of 0.1%. For example, in iron-based superconductors, the strain dependence of the anisotropy in resistivity, that is, the nematic susceptibility, showed a divergence toward the tetragonal-to-orthorhombic transition, which evidenced an electronic nematic phase.[36,37]

On the other hand, applying anisotropic stress is effective for selecting domains from all variants generated by a symmetry-lowering structural transition. In general, when a prototype phase loses part of the symmetry elements, plural orientational states occur in a crystal to form twinning with invariant planes as their interfaces.[49,50] Such a crystal is called a ferroelastic material when identical or enantiomorphous orientational states are freely transferred to each other upon applying mechanical stress.[51,52] Electric field and polarization in ferroelectrics correspond to stress and strain in ferroelastics, respectively; most ferroelastic materials are ferroelectric. Recently, domain control has attracted much attention for realizing novel functionalities by focusing on the properties of the domain interfaces themselves.[53]

A typical example of domain control by adjusting anisotropic stress is found in the high-temperature superconductor $YBa_2Cu_3O_{7-\delta}$. It undergoes a tetragonal-to-orthorhombic transition at 1000 K, resulting in two types of domain connected by {1 1 0} planes. A twin-free crystal was obtained by applying uniaxial mechanical pressures of 50–100 MPa[54-56] and used for the determination of the anisotropy in resistivity in the *ab* plane.[57] This domain control was successful only at high temperatures above 500 K,[56] because the transition was not purely displacive as it is in the martensite transitions but involved a diffusion or a jump of the oxide atoms by ~3 Å;[55] a certain activation energy was necessary for the movement of domain boundaries. In contrast, a martensitic cubic-to-tetragonal transition occurs at $T_m$ above a superconducting transition at $T_c$ in A-15 superconductors such as Nb$_3$Sn ($T_c$ = 18 K, $T_m$ = 36, 43 K) and V$_3$Si ($T_c$ = 17 K, $T_m$ = 21–29 K).[58,59] Resulting domains are identical to those of CRO but with tetragonal distortions of 0.6 and 0.24%, respectively, which are larger by one order of magnitude than that of CRO and smaller than that of YBa$_2$Cu$_3$O$_{7-\delta}$ (3%). A single-domain crystal of V$_3$Si was obtained at low temperatures under a uniaxial pressure of 40 MPa[60] and used for the investigation of strain effects on the superconductivity.[61] Recently, the control of nematic superconductivity domains has been carried out by applying uniaxial stress to the topological superconductor Sr$_x$Bi$_2$Se$_3$.[62]

### 4.1.2 Domain control in CRO

As in the case of the A-15 superconductors, we need a single-domain crystal of CRO to identify the anisotropic properties. It was found in the previous study that domains were controlled to some extent by cooling a crystal in a magnetic field of 7 T,[28] owing to the anisotropy in magnetic susceptibility, which was estimated to be 11% and found to change its sign at $T_{s2}$ (Table 1). However, this magnetic field effect was sample-dependent and thus difficult to use reproducibly. In this study, we applied the anisotropic stress technique using PE elements; since the tetragonal distortion is as small as 0.05%, one expects large stress effects within the experimental limit. In fact, we obtained single-domain crystals with Z and X domains under biaxial and uniaxial stresses giving ~0.05% strains, respectively. The switching of domains at low temperatures was completely reversible, demonstrating that CRO is a ferroelastic compound, similar to the A-15 compounds. Another important finding is a flip of the tetragonal distortion at $T_{s2}$, which is explained in terms of the odd-parity multipolar orders.

### 4.1.3 Structural order parameters

The successive structural transitions of CRO have been discussed in terms of the Landau theory.[32,34] The adopted OPs ($\eta_1$, $\eta_2$), which span the $E_u$ representation of the cubic point group $O_h$, are linear combinations of the displacements of the four Re atoms of one tetrahedron from their ideal positions in phase I, ($x_m$, $y_m$, $z_m$) with $m$ = 1, 2, 3, 4:

$$\eta_1 = (X - Y)/\sqrt{2}, \eta_2 = (X + Y - 2Z)/\sqrt{6},$$

where $X = (x_1 + x_2 - x_3 - x_4)/2$, $Y = (y_1 - y_2 + y_3 - y_4)/2$, and $Z = (z_1 - z_2 - z_3 + z_4)/2$. Phases I, II, and III are characterized by the OPs of ($\eta_1$, $\eta_2$) = (0, 0), (0, $\eta_2$), and ($\eta_1$, 0), respectively. An energy difference between phases II and III finally appears at the eighth-order Landau expansion.[32]

Figure 11 depicts possible deformations of the Re tetrahedron for phases II and III, that is, the three-dimensional crab-hand type and two-dimensional twist type, respectively. For phase II, it is apparent that the Re tetrahedron shown in Fig. 11 expands along the *c* axis. However, no change is expected along *c* in total, because the nearby four tetrahedra connected by their vertices have opposite atomic shifts to shrink along *c*; neighboring tetrahedra are not identical and their shape alternates over the pyrochlore lattice.[6] On the other hand, the in-plane components of the Re shifts also cancel each other. Thus, such a tetragonal distortion as $c > a$ is not straightforwardly expected from the atomic shifts. For phase III, the twist-type shifts cause no change along *c*, while they may reduce *a* because the straight bonds along <1 1 0> become zigzag bonds; neighboring tetrahedra are identical owing to the twofold screw axis.[6] Thus, the resulting deformation is inconsistent with the observed distortion of $c < a$. Therefore, the tetragonal distortions in phases II and III are not directly coupled to these structural OPs. Note, however, that the actual shifts of Re atoms are small, while those of the surrounding oxide atoms are large,[31,43] which may explain the observed tetragonal distortions. In any case, the origin of the tiny tetragonal distortions is unlikely to be a simple structural instability but must be related to an electronic instability.

### 4.1.4 Cluster ETQ orders

The electronic orders of CRO have been considered as odd-parity multipole orders.[1] The cluster-ETQ orders are the most plausible models that are consistent with the symmetry reduction in the crystal structures.[5] Figure 11 gives intuitive descriptions of the cluster ETQs. Let us assume the atomic shifts of Re to be hypothetical electric dipole moments. Then, one could define an electric toroidal (ET) moment at a bond center between two Re atoms. For phase II, ET moments are generated at the four side edges of the tetrahedron, while no ET moment occurs at the top or bottom edge because the electric dipole moments on the bonds cancel each other. As a result, a cluster ETQ of the $x^2 - y^2$ type ($G_v$) appears in the tetrahedron. On the other hand, for phase III, the twist-type atomic shifts generate ET moments only at the top and bottom edges, resulting in a $3z^2 - r^2$-type cluster ETQ ($G_u$). More rigorously, it is proposed that these cluster ETQs are activated by bond or spin-current orders on the Re–Re bonds.[5]

In this cluster ETQ scenario, the true OPs of the successive transitions of CRO are considered to be the magnitudes of ETQ moments. Note that the ETQ moments in the nearby tetrahedra are aligned in the same direction as in the central one in both phases II and III. Thus, the two cluster ETQ orders are ferroic. In general, an electric quadrupole moment can couple with the lattice since it is a biased charge distribution and may cause a uniform lattice distortion, particularly for a ferroic alignment. The resulting lattice distortion was often too small to observe in the case of 4f electron systems,[63] while it was actually observed for more expanded 5d electrons such as in the spin–orbit-coupled insulator Ba$_2$MgRe$_2$O$_6$.[64] Similarly, in CRO with the extended 5d orbitals, the ferroic orders of the ETQ moments must cause reasonably large lattice distortions. Moreover, the observed switching of the tetragonality is likely to occur, because the directions of ETQ moments flip by 90º at $T_{s2}$. This implies that the bonds with ET moments are longer than the other bonds without ET moments in the tetrahedral clusters of ETQ orders.

The cluster ETQ orders of the $x^2 - y^2$ and $3z^2 - r^2$ types belong to the $E_u$ irreducible representation and are degenerate.[5] In an actual compound, however, such degenerate states always take slightly different energies as a result of a small perturbation such as an electron–phonon coupling or a spin–orbit interaction. A similar mutilevel degeneracy is often observed in Jahn–Teller systems with $t_{2g}$ or $e_g$ electrons, in which only one of the states is eventually selected by strong electron–phonon coupling and appears under the ambient condition. For CRO, although the $3z^2 - r^2$-type ETQ is eventually selected as the ground state, the $x^2 - y^2$-type ETQ replaces it upon heating owing to an entropy effect. This is possible because the energy difference between the two phases is minimal as the associated perturbation is small. Therefore, in combination with the tiny tetragonal distortions, it is plausible that the successive transitions of CRO are purely driven by an electronic instability.

*4.2 Anisotropy in electronic states of CRO*
*4.2.1 Anisotropy in resistivity and Fermi surfaces*

The anisotropy in resistivity measured on the single-X-domain crystal was significantly large in spite of the small tetragonal distortions: it grows below $T_{s1}$, reaches 25% ($\rho_c > \rho_a$), and changes to –25% ($\rho_c < \rho_a$) immediately below $T_{s2}$. In general, resistivity is given by $m^*/ne^2\tau$, in which $m^*$ and $n$ are the effective mass and the density of carriers, and $e$ and $\tau$ are the elemental charge and scattering time, respectively. Thus, the anisotropy in resistivity is determined by the shape of FSs and anisotropy associated with scattering mechanism. Although a tetragonal crystal naturally has a certain anisotropy, our band structure calculations revealed that simple deformations of FSs of phase I by a tetragonal distortion of 0.05% are too small to explain the observed anisotropies.

According to the previous band structure calculations, phase I is a compensated metal with two spherical electron-like FSs around the Γ point and one hole-like FS near the K point.[65,66] It is expected that each of the FSs will split into a pair of FSs with spin–momentum locking in phases II and III as a result of the activation of ASOC by ISB. Thus, three pairs of spin-split FSs should appear in total. In fact, the previous calculation for phase III revealed six FSs, which were mostly consistent with the de Haas–van Alphen effect measurements by means of magnetic torque.[29] The estimated magnitude of ASOC is 67 K, which is much smaller than 4600 K for the Rashba metal BiTeI,[67] because of the small polar electric field induced by the structural distortions for CRO compared with the large one for BiTeI originating from the coordination of the different anions around the Bi atom in its built-in structure. It is emphasized that this subtle nature in symmetry breaking caused by the phase transitions is a unique characteristic of CRO. Our band structure calculations revealed the three pairs of spin-split FSs for either phase II or III, as shown in Appendix Figs. A1 and A2, respectively. Note that, as a result of the splitting, almost half of the hole FS is removed with its energy being above the Fermi energy in each phase, which is the main reason for the large loss in the density of states below $T_{s1}$.

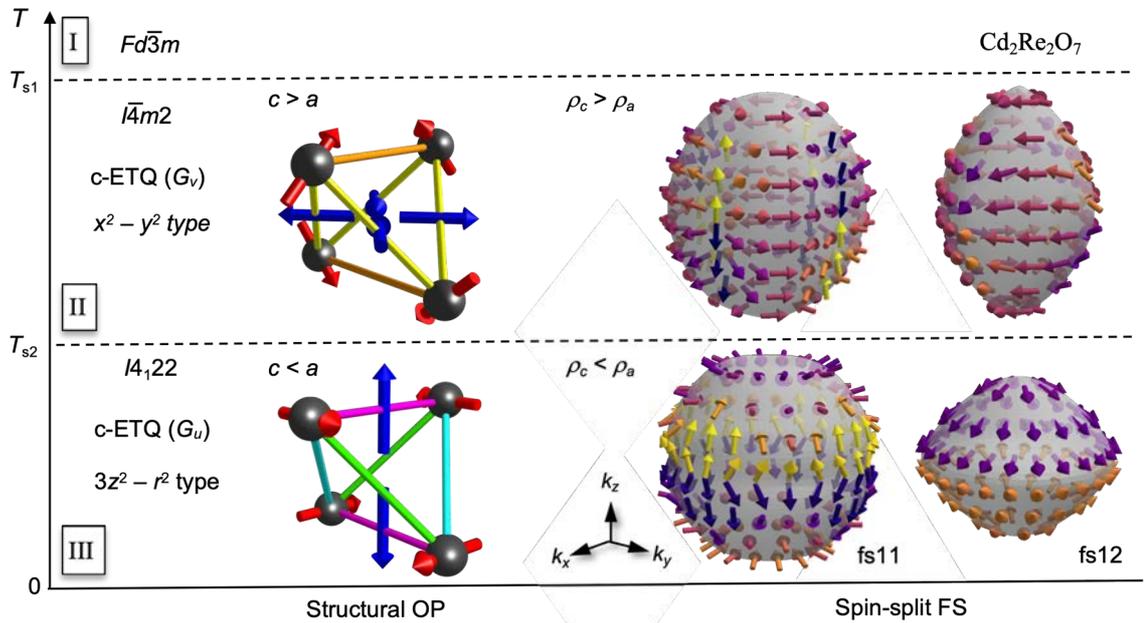

**Fig. 11.** (Color online) Temperature evolution of the structural OP and the pair of spin-split FSs derived from the smaller electron FS of phase I; outer (fs11) and inner FSs (fs12) are depicted on the left and right, respectively; pseudocubic cells are assumed. For each structural OP, a set of red arrows at the four Re atoms (black balls) in the tetrahedral cluster show the atomic shifts allowed in the corresponding space group.[32,34] The colors of bonds distinguish identical bonds in the deformed tetrahedra. The blue arrow at the center of each bond represents an electric toroidal moment generated by the pair of electric dipole moments (Re shifts) at the two ends of the bond. For phase II, the four electric toroidal moments on the side bonds form a cluster electric toroidal quadrupole (c-ETQ) of the $x^2 - y^2$ type ($G_v$), while the two on the top and bottom bonds form a $3z^2 - r^2$ type c-ETQ ($G_u$) for phase III.[5] The arrows on the FSs represent the directions of spins in the two types of spin-momentum locking induced by the ASOCs for phases II and III. Their colors distinguish the magnitudes of the $z$ components of spin: typically, positive (yellow to orange), nearly zero (red), and negative (violet to blue).

The electron FSs must play a major role in determining the transport properties, because the holes are relatively heavy[65] and have a lower mobility, as indicated by the observations of negative Hall coefficients[10,68,69] and negative Seebeck coefficients.[46] In this section, we consider the deformations of the FSs, which cause anisotropies in the effective mass. Then, in the next section, we discuss the effect of carrier scattering. The observed anisotropies can be qualitatively explained when a spherical FS in phase I elongates along the $k_z$ direction in phase II ($m^*_c > m^*_a$) and shrinks in phase III ($m^*_c < m^*_a$). To obtain ±25% anisotropies as observed, spheroidal FSs with axial ratios of ±25% are to be assumed. In fact, the FSs calculated for phase III flatten along $k_z$.[29] Our band structure calculations confirmed this flattening for phase III in addition to the elongation of the FSs for phase II, as illustrated in Fig. 11 for the pair of spin-split FSs derived from the smallest electron FS of phase I. Therefore, the deformations of the FSs are in line with the observed anisotropies in resistivity.

Using the calculated band structures, we have calculated the anisotropy in the effective mass $m^*_c/m^*_a$ from the ratio of projected areas along the $k_z$ and $k_x$ directions. For the inner FSs shown in Fig. 11, they are 22 and –30% for phases II and III, respectively, which are in good agreement with 25 and –25% obtained in experiments, respectively. However, combining the anisotropy for the outer FSs of Fig. 11 gives average values of 22 and –7%, respectively, with the latter being considerably reduced. Moreover, averaging over the two pairs of the electron FSs gives –0.2 and –6% for phases II and III, respectively. Finally, including all the three pairs of FSs gives –7 and 4%, respectively, the signs of which are opposite those of the experimental values. Thus, the observed anisotropies in resistivity are explained only for the inner FSs, and it is doubtful whether only the deformations of FSs cause such large anisotropies in resistivity as observed.

For more qualitative examinations, we have calculated anisotropy in conductivity via Boltzmann's transport equation based on the calculated FSs, assuming that the scattering rates are isotropic. The values of $2(R_c - R_a)/(R_c + R_a)$ are estimated in the low-temperature limit to be 6 and –8% for phases II and III, respectively. The signs of these values are consistent with the results of experiments, possibly because minor contributions from relatively heavy holes have been removed. Moreover, the value of phase III is close to the observed lowest-temperature value of –10% in Fig. 10. However, they are merely less than half the maximum values at around $T_{s2}$. Therefore, considering the deformations of FSs is adequate at low temperatures but not at elevated temperatures, and we must take into account the role of quasiparticle scattering.

*4.2.2 Quasiparticle scattering in CRO*

The marked change in the temperature dependence of resistivity at $T_{s1}$ indicates that a crucial change takes place in quasiparticle scattering. The temperature dependence of resistivity is almost flat above $T_{s1}$ up to 300 K in Fig. 9, as it was in the previous studies up to 500[70] and 600 K[46] with large values of 300–800 μΩ cm.[2,8,20,71,72] This "saturated" resistivity must be interpreted in terms of the Mott–Ioffe–Regel limit.[73] For A-15 superconductors, for example, the mean free path of carriers becomes small, comparable to the interatomic distance, owing to strong electron–phonon scattering, giving rise to such saturation. By analogy, the sudden decrease in resistivity in CRO means that strong scattering (probably by electron–phonon couplings) active in phase I has been removed below $T_{s1}$ because of the ISB and the accompanying spin splitting of the FSs.

We will emphasize here the important role of spin-dependent scattering in the spin-split FSs, which has been well established in Rashba metals without inversion symmetry and with large SOCs. For example, in BiTeBr with a gigantic Rashba splitting as in BiTeI,[67] the interference of quasiparticles depends on the spin texture of the split FSs:[74]

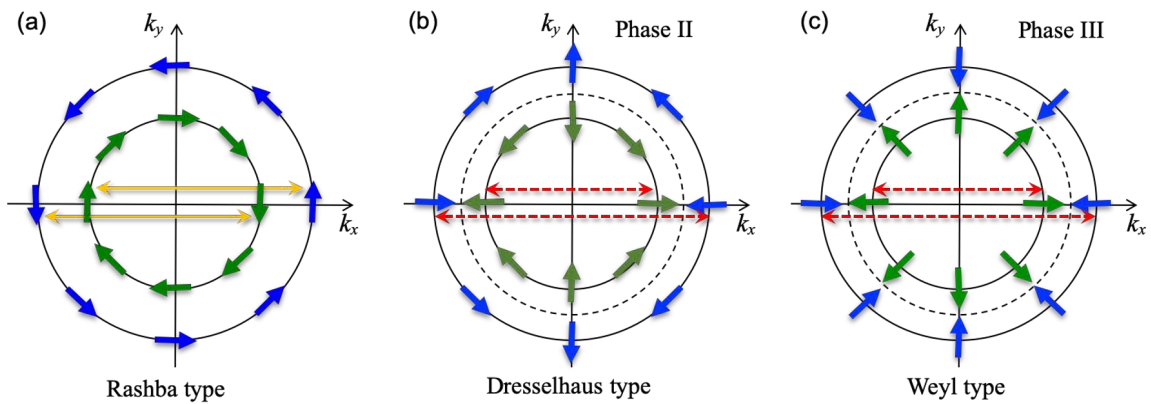

**Fig. 12.** (Color online) Three types of spin-momentum locking in the $k_x$–$k_y$ plane for metallic systems without inversion symmetry and with large spin–orbit interactions: (a) Rashba type with point group $C_{4v}$, (b) Dresselhaus type with $D_{2d}^{[110]}$ corresponding to phase II, and (c) Weyl type with $D_4$ corresponding to phase III. The blue and green arrows represent the directions of spin in the outer and inner FSs, respectively. The spin textures in (b) and (c) correspond to the case of $c_1 < 0$ in the ASOC forms described in the text; for each spin texture, there is a counterpart with a different chirality ($c_1 > 0$) that is connected by space inversion symmetry. The yellow arrows in (a) represent dominant backscattering processes conserving the quasiparticle spin, and the dotted red arrows in (b) and (c) represent scattering vectors between points on the opposite sides of the same FSs, which are strongly suppressed as they require spin flipping. The dotted circles in (b) and (c) show the spin-degenerate FS of phase I with inversion symmetry.

as depicted in Fig. 12(a), backscattering between inner and outer FSs can occur as it is spin-conserving, while those within the same FSs are not allowed as they require spin flipping.

Similar spin-dependent scattering effects are expected for CRO, although the types of ASOC for CRO are different from those of the Rashba metal with the point group $C_{4v}$. According to the symmetry analyses by Hayami et al.,[5] the ASOC contributions in CRO are given as $c_1(k_x\sigma_x - k_y\sigma_y)$ for phase II with the point group $D_{2d}^{[110]}$ and $c_1(k_x\sigma_x + k_y\sigma_y) + c_2 k_z\sigma_z$ for phase III with the point group $D_4$, ignoring higher order terms; $c_1$ and $c_2$ are coefficients proportional to the OPs, $\sigma$ is the Pauli matrix, and a pseudocubic lattice is assumed. The former gives a Dresselhaus-type ASOC having only in-plane spin components with rotations different from those of the Rashba type [Fig. 12(b)]. The latter has three-dimensional spin components with the in-plane ones of the Weyl-type all-in–all-out configurations [Fig. 12(c)]. In either case, backscattering within the same FSs is not allowed as it requires spin flipping, as in the Rashba metal. Thus, one expects that the large phonon scattering present in the spin-degenerate FSs of phase I will be partly suppressed as spin splitting develops below $T_{s1}$. This must be one of the reasons for the observed reduction in resistivity. It is interesting to note that, in spite of the time reversal symmetry not being broken, spin-dependent scattering can dominate transport properties in such spin-split FSs induced by space ISB.

Another factor causing a reduction in resistivity below $T_{s1}$ is an increase in carrier density via the band Jahn–Teller mechanism, as proposed previously.[68] Our band structure calculations estimate the carrier densities as $n_h = n_e = 4.9 \times 10^{20}$, $7.6 \times 10^{20}$, and $7.7 \times 10^{20}$ cm$^{-3}$ for phases I, II, and III, respectively; the value for phase III is in good agreement with the experimental value of $8.1 \times 10^{20}$ cm$^{-3}$.[68] Thus, the values for the LT phases increase by 60% compared with that for phase I. Since the carrier density gradually increases upon cooling to below $T_{s1}$ in the second-order manner, it causes the observed reduction in resistivity. Nevertheless, it cannot explain the anisotropy in resistivity. Therefore, the spin-dependent scattering must play an important role in determining the transport properties.

*4.2.3 Anisotropies in spin-dependent scattering*

The observed anisotropies in the resistivity of phases II and III must be caused by spin-dependent scattering in the spin-split FSs. The calculated spin textures are shown in Fig. 11 for the pair of FSs derived from the smallest FS of phase I (fs11 and fs12); the other FSs basically have similar spin textures, as shown in Appendix Figs. A1 and A2. The spin textures are mostly similar to those expected from the ASOC expressions mentioned previously (Fig. 12), but are different in parts because of the higher-order terms in the ASOCs and the complex FSs. For phase II, most electrons have only in-plane spin components, whereas, for phase III, the $\sigma_z$ components are relatively large and change their signs across $k_z = 0$. Carrier scattering in conduction along the $c$ axis can occur easily in phase II as it is mostly spin-conserving, whereas in phase III it is suppressed as the $\sigma_z$ components must be flipped with the in-plane components conserved. Thus, $\rho_c$(II) should be larger than $\rho_c$(III), which is consistent with the change at $T_{s2}$ in Fig. 9. On the other hand, carrier scattering in conduction along the $a$ axis must be suppressed in both phases as it always requires the flipping of the in-plane spin components. However, the suppression must be larger in phase II than in phase III, because the in-plane components are relatively large in phase II. Thus, one obtains the relation $\rho_a$(II) < $\rho_a$(III), which is consistent with the change at $T_{s2}$ in Fig. 9. As a result, the origin of anisotropy is clearly explained by the anisotropic spin-dependent scattering. Therefore, the observed anisotropy and its change at $T_{s2}$ reflect the unique characteristics of the spin–momentum locking realized in CRO.

The factors that scatter carriers effectively are likely soft-phonon modes associated with the structural transitions in CRO, which were actually observed by Raman scattering experiments.[33] Such carrier scattering by them must be critical especially near the transition temperatures. The reason why the anisotropy in resistivity is maximum at around $T_{s2}$ is that the spin-dependent scattering effects become more pronounced owing to the enhanced electron–phonon couplings by the soft phonons associated with the $T_{s2}$ transition. In addition, the origin of the strong scattering above $T_{s1}$ leading to the Mott–Ioffe–Regel limit may be electron–phonon scattering by soft phonon modes towards the $T_{s1}$ transition.

At low temperatures, the spin-dependent backscattering becomes less active, and conventional forward scattering by long-wavelength acoustic phonons, which is always allowed as it conserves spin, should dominate the scattering process. Thus, one expects no particular anisotropy based on spin-dependent scattering. In fact, as mentioned previously, the anisotropy of –10% for phase III at the lowest temperature is mostly explained by the deformation of the FSs. Nevertheless, there is a possibility that acoustic phonons play some role. Generally, for systems without inversion symmetry, acoustic phonons can have electric polarizations and may cause strong polar scattering. This is called the piezoelectric scattering and has been observed in zincblende and wurtzite semiconductors such as GaAs and CdS.[75] The piezoelectric scattering can have a certain anisotropy; for CdS, it causes anisotropy to increase by a factor of two in the mobility. For CRO, such piezoelectric scattering may contribute to anisotropy to some extent at low temperatures, although its effect on anisotropy may be small owing to the small tetragonal distortions. Further experimental study on the phonon modes and theoretical considerations on the scattering mechanism are required to discuss the transport properties of CRO in more detail.

*4.2.4 Anisotropy in other properties and future prospects*

Not only anisotropy in resistivity but also other anisotropic properties can be related to the spin-split FSs in CRO. The observed anisotropy in magnetic susceptibility and its turnover at $T_{s2}$ (Table 1) is one example. Since the time reversal symmetry is not broken, it is unlikely that the spin part of the magnetic susceptibility $\chi_s$ causes the anisotropy. The previous Re NQR experiments revealed a large orbital part $\chi_{orb}$, which was $3.16 \times 10^{-4}$ cm$^3$ mol$^{-1}$ and happened to be equal to the $\chi_s$ part.[14] The large $\chi_{orb}$ contribution was suggested to be due to orbital fluctuations of the Re 5d electrons.[14] It is not clear, but there must be a specific mechanism by which the $\chi_{\rm orb}$ causes the observed anisotropy in magnetic susceptibility.

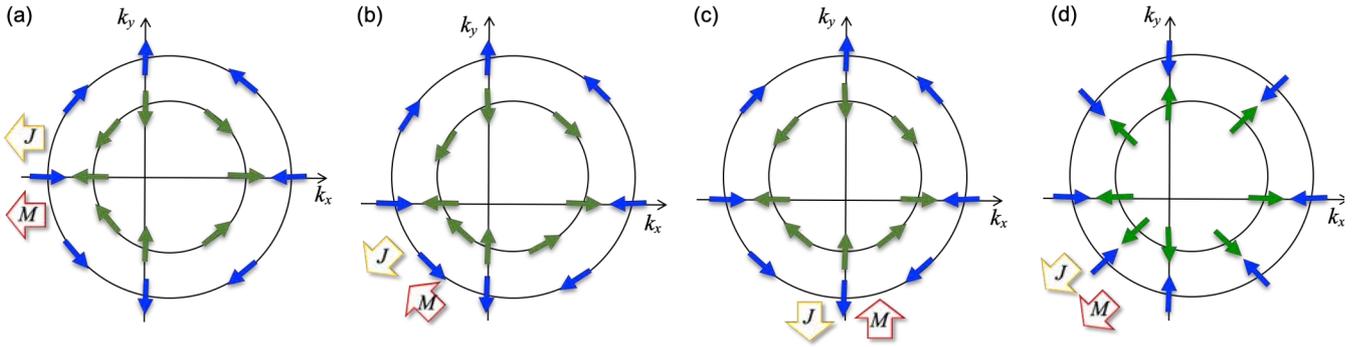

**Fig. 13.** (Color online) Schematic representations of the magneto-current effects predicted for phases II (a–c) and III (d). The pair of spin-split FSs in Fig. 12(b) for phase II shifts as depicted when an electric current $J$ is applied along $-x$ (a), $[-1\ -1\ 0]$ (b), or $-y$ (c). For example, in (a), the outer and inner FSs shift to the right so that they have uncompensated magnetic moments along $-x$ and $x$, respectively. Since the number of spins is larger in the outer FS than in the inner FS, a net magnetization $M$ is generated along $-x$ in the same direction as the current. In similar fashions, magnetizations perpendicular and antiparallel to the current directions are induced in (b) and (c), respectively. In (d) for phase III, a magnetization is isotropically generated in the same direction as the current.

The most intriguing topics to be explored in the future are the complex and anisotropic off-diagonal responses to the application of electric current owing to magneto-current effects.[5] For example, Fig. 13 illustrates how the spin-split FSs should respond to electric currents $J$ applied along different directions in the $x$–$y$ plane. For phase II, when $J$ is along $-x$, the centers of gravity of both the outer and inner FSs shift along $x$, so that they have uncompensated magnetic moments along $-x$ and $x$, respectively [Fig. 13(a)]. As a result, a net magnetization $M$ should appear along $-x$ because of the larger number of electrons in the outer FS; the directions of $J$ and $M$ are the same. In similar manners, $M$ along $y$ occurs in the opposite direction to $J$ along $-y$ [Fig. 13(c)], and transverse $M$ is generated for diagonal $J$ [Fig. 13(b)]. On the other hand, an isotropic response is expected for phase III: $M$ always occurs in the same direction as $J$ [Fig. 13(d)]. It would be interesting to examine these complex off-diagonal responses and compare them with theoretical expectations quantitatively. Note, however, that one expects more complicated responses from the actual spin-split FSs. In addition to the magneto-current effects, a nonreciprocal transport in an applied magnetic field and a lattice distortion induced by a current in a magnetic field are proposed to occur in CRO.[5]

The last topic is the presence of a pair of chirality domains inside one tetragonal domain in each of phase II or III, where the chirality domains are related to each other by space inversion. In fact, Harter et al. observed them in their nonlinear optics experiments.[26] The counter chirality domain for each phase has a spin texture obtained by flipping all the spins in Fig. 12; this spin texture is derived from a coefficient of the opposite sign in the ASOC term: $c_1 < 0$ for Fig. 12 and $c_1 > 0$ for the counterpart. This chirality must couple via the ASOC with the structural phase in the alternation of elongated and compressed tetrahedra in phase II or with the rotation chirality arising from the twofold screw axis in phase III. In a racemic mixture of even numbers of enantiomers, any off-diagonal response is averaged, even in a "single-domain" crystal. However, unbalance between such chirality pairs often occurs in nature, so that there is a chance to observe the responses. Otherwise, we will attempt to obtain a true single-domain crystal by combining electrical current (much larger than that used in the present resistivity measurements), magnetic field, and mechanical strain all at once.

## 5. Conclusions

We have precisely controlled the tetragonal domains of the LT phases of the SOCM candidate CRO under biaxial and uniaxial stresses. The POM observations on the (0 0 1) surface revealed that single-Z-domain crystals were obtained under compressive and expansive biaxial stresses in the plane for phases II and III, respectively. Moreover, single-X-domain crystals were obtained under expansive and compressive uniaxial stresses along the $x$ direction for phases II and III, respectively. These variations are completely reversible as a function of strain, demonstrating that CRO is a ferroelastic material. From the observed responses of domains to stresses, we concluded that the tetragonal distortion is flipped between $c > a$ and $c < a$ for phases II and III, respectively. The anisotropy in resistivity was measured using single-X-domain crystals, and it was found that the anisotropy is as large as 25% and flips at $T_{s2}$. The origin of the anisotropic transport was ascribed to the spin-dependent scattering in the spin-split FSs as well as the deformation of the FSs.

Our findings are basically consistent with the $E_u$ OPs leading to odd-parity multipolar orders described by cluster-ETQs.[5] However, recent nonlinear optics and magnetic torque measurements suggested contributions of other OPs such as $T_{2u}$, $T_{1g}$, and $E_g$,[23,24,26,27] which may be related to more complex electronic orders. It seems that very rich physics is involved in CRO and thus in the SOCMs awaiting further investigation.

**Acknowledgments**

The authors are grateful to Y. Motome, M. Takigawa, M. Ogata, and T. Hasegawa for helpful comments. This work is based on S.T.'s master thesis written during fiscal years 2019 and 2020. This work was supported by KAKENHI (Grants No. 18K13491, 18H04308, and 18H04323) from Japan

Society for the Promotion of Science (JSPS). It was also supported by a Core-to-Core Program [(A) Advanced Research Networks] and a Grant-in-Aid for Scientific Research on Innovative Areas "Quantum Liquid Crystals" (Grant No. 20H05150) from JSPS.

**Author contributions**

S.T., D.H., Y.K., and M.T. performed the POM observations. D.H. prepared the crystals. S.T., K.A., and T.C.K. performed the resistivity measurements. H.T.H calculated the band structures. S.T., D.H., and Z.H. carried out data analyses. D.H. and Z.H. conceived and designed the project. All authors contributed to manuscript preparation and discussion.

**Appendix**

Three calculated pairs of FSs are shown in Fig. A1 for phase II and in Fig. A2 for phase III.

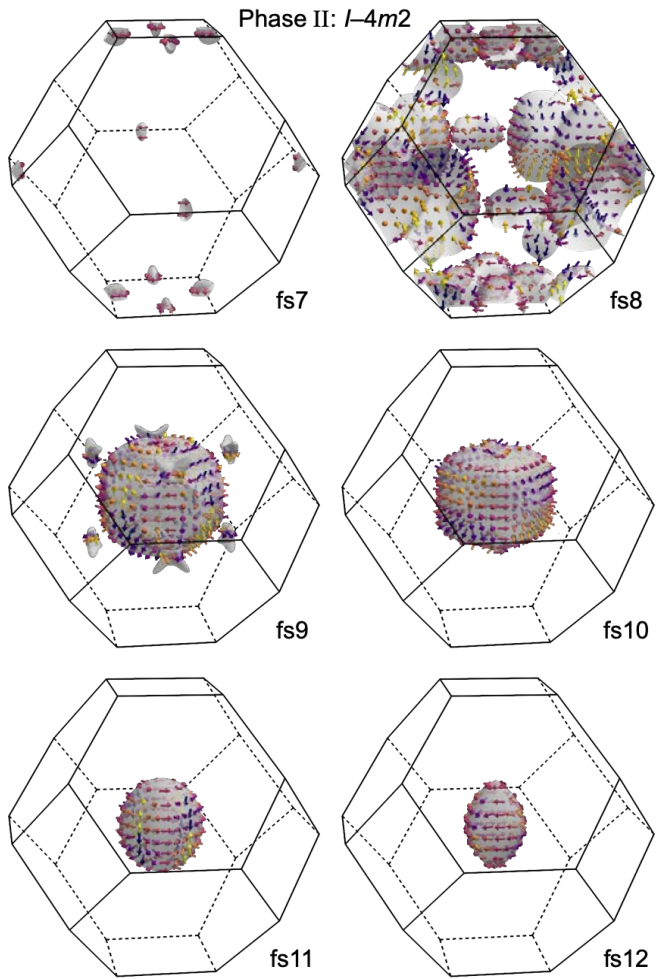

**Fig. A 1.** (Color online) Calculated FSs for phase II of CRO. fs7 and fs8 are the pair of spin-split hole FSs derived from the spin-degenerate hole FS in phase I. fs9–fs10 and fs11–fs12 are the pairs of outer (left) and inner (right) spin-split electron FSs derived from the two spin-degenerate electron FSs in phase I. The arrows on each FS represent the direction of spins and their colors indicate the magnitudes of the $z$ components of spin: typically, positive (yellow to orange), nearly zero (red), and negative (violet to blue).

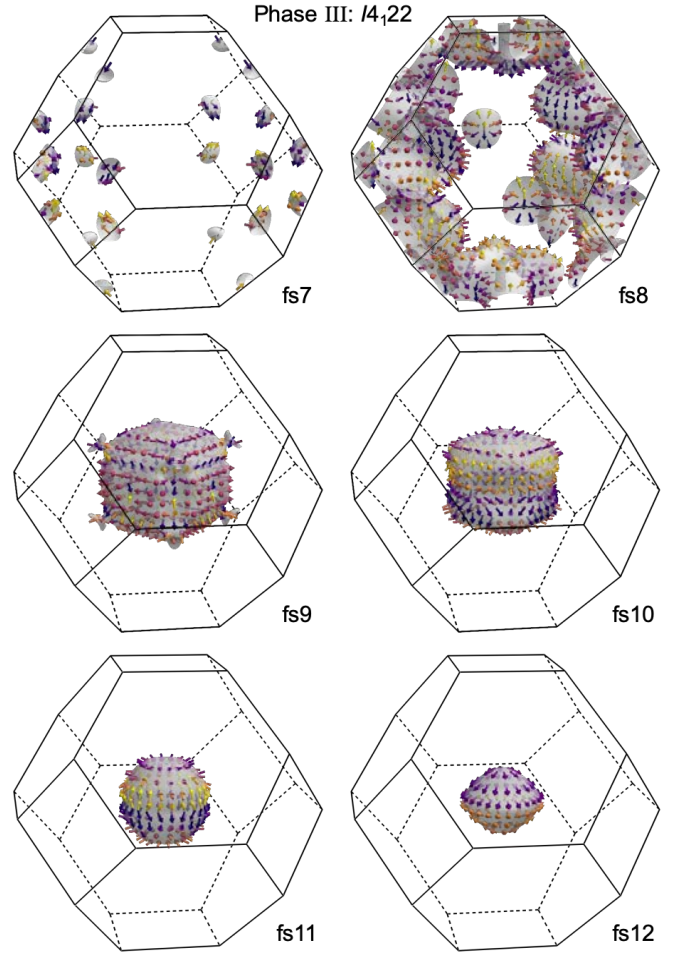

**Fig. A 2.** (Color online) Calculated FSs for phase III of CRO. fs7 and fs8 are the pair of spin-split hole FSs derived from the spin-degenerate hole FS in phase I. fs9–fs10 and fs11–fs12 are the pairs of outer (left) and inner (right) spin-split electron FSs derived from the two spin-degenerate electron FSs in phase I. The arrows on each FS represent the direction of spins and their colors indicate the magnitudes of the $z$ components of spin: typically, positive (yellow to orange), nearly zero (red), and negative (violet to blue).